\documentclass{article}

\usepackage[utf8]{inputenc} 
\usepackage[T1]{fontenc}    
\usepackage[hidelinks]{hyperref}       
\usepackage{url}            
\usepackage{booktabs}       
\usepackage{amsfonts}       
\usepackage{nicefrac}       
\usepackage{microtype}      
\usepackage{lipsum}
\usepackage{tikz}
\usepackage{dsfont}
\usepackage{amsmath}
\usepackage{array}
\usepackage{todonotes}
\usepackage{float}
\usepackage{rotating}
\usepackage[toc,page]{appendix} 
\usepackage{authblk}
\usepackage{geometry}
\usepackage{placeins}
\usepackage[numbers]{natbib}

\renewcommand{\figurename}{Fig.{}}

\usepackage{color,soul}

\usepackage{xpatch}
\xpatchcmd{\author}{\relax#1\relax}{\relax\detokenize{#1}\relax}{}{}

\title{Bayesian ensemble modelling to monitor excess deaths during summer 2022 in Switzerland}

\author[1,*]{Garyfallos Konstantinoudis}
\author[2]{Anthony Hauser}
\author[3]{Julien Riou}

\affil[1]{MRC Centre for Environment and Health, Department of Epidemiology and Biostatistics, School of Public Health, Imperial College London, London, UK}
\affil[2]{INSERM, Sorbonne Universit{\'e}, Pierre Louis Institute of Epidemiology and Public Health, Paris, France}
\affil[3]{Institute of Social and Preventive Medicine, University of Bern, Bern, Switzerland}

\affil[*]{Corresponding author, email: \url{g.konstantinoudis@imperial.ac.uk}}
\date{}                    

\newcolumntype{L}[1]{>{\raggedright\let\newline\\\arraybackslash\hspace{0pt}}m{#1}}
\newcolumntype{C}[1]{>{\centering\let\newline\\\arraybackslash\hspace{0pt}}m{#1}}
\newcolumntype{R}[1]{>{\raggedleft\let\newline\\\arraybackslash\hspace{0pt}}m{#1}}

\begin{document}
	
	\maketitle
	
	
	\begin{abstract}
		Switzerland experienced one of the warmest summers during 2022. Extreme heat has been linked to increased mortality. Monitoring the mortality burden attributable to extreme heat is crucial to inform policies, such as heat warnings, and prevent heat-related deaths. In this study, we evaluate excess mortality during summer 2022, identify vulnerable populations and estimate temperature thresholds for heat warnings. We use nationwide mortality and population data in Switzerland during 2011-2022 by age, sex, day and canton. We develop a Bayesian ensemble modelling approach with dynamic population to predict expected mortality in summer 2022 and calculate excess by comparing expected with observed mortality. We account for covariates associated with mortality such as ambient temperature, national holidays and spatiotemporal random effects to improve predictions. After accounting for the effect of the COVID-19 pandemic, we found a 3\% (95\% credible interval: 0\%-6\%) excess mortality during summer 2022. We observed a total of 456 (5-891) excess deaths during summer 2022 in people older than 80 years. There was weak evidence of excess mortality in the other age groups. The highest excess mortality was observed in July (12\%: 4\%-19\%), the hottest month in 2022. We also found that for heatwaves longer than four days, the minimum excess mortality temperature threshold in the oldest age group is the 70th percentile of the temperature. To reduce future summer excess mortality in Switzerland, we propose targeted heat warnings to older populations and reducing the temperature threshold when weather forecasts predict periods of extreme heat of four days or longer.
	\end{abstract}
	
	\section{Introduction}
	In 2022, Switzerland experienced the second warmest summer ever after the extreme summer of 2003 \cite{Copernicus,vicedo2023footprint}. Extreme heat is responsible for a substantial health burden, including premature deaths and hospital admissions, with frail populations (older people with chronic conditions) being affected the most \cite{konstantinoudis2022ambient, konstantinoudis2023asthma, masselot2023excess, van20222022}. Quantifying the mortality burden attributable to the extreme heat during 2022 is very important for informing policies to protect vulnerable populations but challenging \cite{de2022assessing, riou2023direct, van2022covid, msemburi2023estimates}. The emergence of the omicron BA-4 and BA-5 variants of SARS-CoV-2 led to a wave of infections that was concurrent with the heatwaves, making it difficult to disentangle the consequences of heat and SARS-CoV-2. In addition, the COVID-19 pandemic disrupted mortality patterns and population trends in 2020-2021, imposing further challenges when estimating mortality attributable to heat in 2022 \cite{riou2023direct}. 
	
	Several studies have quantified the effect of extreme heat in summer 2022 \citep{ vicedo2023footprint, Ballester2023, raglettiMonitoring2023, pascal2023estimation}. Two studies in Switzerland estimated the effect of temperature on all-cause mortality using historical data and calculated the number of deaths attributable to heat exposure \citep{vicedo2023footprint, raglettiMonitoring2023}. A study in different European countries using similar methodology estimated temperature-related deaths at the sub-regional level during summer 2022 \citep{Ballester2023}. A study in France, while controlling for a possible influence of the COVID-19 pandemic, estimated deaths attributable to heat and heatwaves during 2014-2022 using daily temperature in 96 French departments \citep{pascal2023estimation}.
	
	The above approaches are complementary and important tools for mortality surveillance due to heat exposure, but they have important limitations. 
	The outputs are {\em expected} mortality due to heat exposure based on historical rather than current data. These estimates are hard to validate and rely on strong assumptions regarding causality (e.g., accounting or not for the role of air-pollution in the causal effect of extreme heat on mortality\cite{buckley2014commentary}), vulnerable populations (e.g., the effect varies by age, sex and in space \cite{konstantinoudis2022ambient}) and temporal adaptation (e.g., periods during which the effect of heat exposure is assumed to be the same\citep{vicedo2018multi}).  An alternative way of assessing the impact of extreme heat is by estimating excess all-cause mortality, i.e. comparing the observed mortality with the counterfactual scenario that the extreme event did not occur. This counterfactual can be obtained by extrapolating from historical data on mortality while accounting for factors affecting all-cause mortality such as population ageing, ambient temperature and public holidays \citep{blangiardo2020estimating, konstantinoudis2022regional, kontis2020magnitude, riou2023direct}. This methodology has several advantages over the above, including fewer assumptions, direct ways to account for concurrent events (e.g., COVID-19 waves), and a natural way to propagate uncertainty from all sources thanks to the Bayesian framework.
	
	In this study, we proposed a Bayesian ensemble approach to estimate excess mortality during the extreme heat periods of summer 2022 in Switzerland. Our novel approach accounts for COVID-19 mortality, mortality trends across the different age-sex groups, spatial auto-correlation and long- and short-term  population changes. We did not make any assumptions about the shape and magnitude of the effect of extreme temperature on all cause-mortality. As excess mortality is estimated daily at the local level by age and sex, our approach supports aggregating the estimated excess deaths by any union/intersection of the temporal-, spatial-, sex- and age-specific groups desired. Last, we used this  approach to estimate temperature thresholds to inform heat-warnings and to understand the interplay between extreme heat and COVID-19 mortality.

	\section{Methods}
	
	We obtained all-cause mortality data at daily temporal and cantonal (level 3 in the nomenclature of territorial units for statistics \cite{NUTSregions}) spatial resolution during 2011-2022 in Switzerland from the Swiss Federal Office of Public Health. Data was aggregated by age ($<40$, 40-59, 60-69, 70-79 and $>80$) and sex (males-females). The selection of the age subgroups was in line with the literature, \cite{davies2021community, konstantinoudis2022regional}. Individual data on laboratory-confirmed COVID-19 deaths during 2020-2022 was available from the Swiss Federal Office of Public Health. Data included information about age, sex, canton of residence and the date and type of the positive SARS-CoV-2 test. Age-, sex- and canton-specific population size was available for the 31st of December of the years between 2010 and 2021 from the Federal Statistical Office in Switzerland. We used temperature and national holidays to improve the predictive ability of the models. Daily mean ambient temperature between 2011 and 2022 at 0.25$^o \times$0.25$^o$ resolution was retrieved from the ERA-5 reanalysis data \citep{hersbach2020era5}. Holidays were obtained for each canton (1 if there is a cantonal holiday, 0 otherwise). The cantonal mean temperature was calculated using municipality-level population weights within each canton \cite{de2021comparative}. 
	
	\subsection*{Population model}
	We predicted daily population counts by age, sex and canton following a two step procedure. On step one we fitted and averaged with uniform weights 4 Bayesian hierarchical spatiotemporal models. Let $P_{gst}$ be the population of the $g$-th age and sex group, in the $s$-th canton in the 31st of December of the $t$-th year. We fitted the following models for each $g$-th group separately. For simplicity we omit $g$ from the notation in this subsection:
	\begin{align*}
		P_{st} &\sim \text{Poisson}(\lambda_{st}) \\
		\log(\lambda_{st}) &= \alpha_{0} + \epsilon_{k} + u_{s} + v_{st} + b_{*} \\
		\epsilon_{k}, u_{s}, v_{st}|\tau_{p} &\sim \text{Normal}(0, \tau_{p}^{-1}), \text{\ with } p=\{1,2,3\}
	\end{align*}
	\noindent where $\alpha_{0} + \epsilon_{k} + u_{s} + v_{st}$ is a random intercept per k-th language region (French, German and Italian), s-th canton, st-th spatiotemporal unit and $b_*$ a term to capture the temporal trends defining the 4 different models, for model 1: $b_t = \alpha_1 t$, for model 2: $b_t=w_t$ defining a random walk of order 1, $w_t|w_{t-1}, \tau_{4} \sim \text{Normal}(w_{t-1}, \tau_{4}^{-1})$, for model 3: $b_t= \alpha_1 t + w_t$ and model 4 defines a random slope by canton: $b_{st}= \alpha_1 t + r_t + \alpha_{s t}$, where $r_t, \alpha_s \sim \text{Normal}(0, \tau_{p}^{-1}), \; p=5,6$. The second step was to calculate daily population. Let $O_{lts}$ be the observed number of deaths on day $l$ of year $t$ and canton $s$, then:
	$$P_{st(l+1)} = P_{stl} - O_{stl} + c_{st}$$
	where $c_{st}$ is the is the net change in population size due to births, ageing and migration, assumed to be constant throughout the year, with $c_{st} = (P_{s(t+1)} - P_{st} - O_{st})/365.25$, with $P_{st(l+1)}$ being data input for the years 2011-2021 and a random variable for the year 2022. We refer to the daily population size as $\Tilde{P}_{stl\xi}$, where $\Tilde{P}_{stl\xi} = P_{stl}$, for $t<2022$ and $\Tilde{P}_{stl} = P_{stl\xi}$, for $t=2022$ and the $\xi$-th sample of the posterior predictive.
	
	\subsection*{Expected mortality model}
	To estimate the expected mortality in summer 2022, we followed a similar Bayesian ensemble approach as for the population. Let $O_{stlm}$ be the number of all-cause deaths and $\Tilde{P}_{stlm\xi}$ the population in the $s$-th canton on the $l$-th day of the $m$-th week of the $t$-th year. Note that as for the population model we fitted the models separately for each g-th group and we omit from the notation for simplicity:
	\begin{align*}
		O_{stlm} &\sim \text{Poisson}\big(\mu_{stlm}\Tilde{P}_{stlm\xi}\big) \\
		\log(\mu_{stlm}) &= \beta_{0} + \beta_1 X_{1sl} + \beta_2 X_{2sl} +  f(r_{stl}) + z_{s} + q_{tm} + \gamma_{*}
	\end{align*}
	\noindent where $\beta_0$ is an intercept term, $\beta_1$ the effect of holidays, $\beta_2$ the effect of COVID-19 ($X_{2sl}=1$ if there is at least one COVID-19 death in the $s$-th canton, $l$-th day and $g$-th group and $X_{2sl}=0$ otherwise; this is because there was just 34 days-cantons with more than 1 COVID-19 deaths) $f(r_{stl})$ captures the non-linear effects of temperature, $z_{s}$ spatial autocorrelation, $q_{tm}$ seasonal summer effects and $\gamma_{*}$ the long term temporal trend, Supporting Information Appendix Table SI17. The non-linear effect $f(\cdot)$ of the mean daily temperature in each canton, $r_{stl}$ are defined using a second order random walk process $r_{stl}|r_{st(l-1)}, r_{st(l-2)}, \tau_7 \sim \text{Normal}\big(2r_{st(l-1)} + r_{st(l-2)}, \tau_7^{-1}\big)$. The spatial autocorrelation is modelled using an extension of the Besag York Molli{\'e} (BYM) model, writing $z_s = 1/\sqrt{\tau_8} (\sqrt{1-\phi}d_{1s} + \sqrt{\phi}d_{2s})$, where $d_{1s} \sim \text{Normal}(0,1)$ and $d_{2s}$ is the intrinsic conditional autoregressive prior standardised so the marginal variance is equal to 1 \cite{besag1991bayesian, rieblerbym2, konstantinoudis2020discrete, simpsonpenal}. The term $\tau_8$ is the precision, whereas $\phi$ a mixing parameter. We model seasonality through an autoregressive prior of order 1 at the weekly level, $q_{tm}|q_{t(m-1)}, \rho, \tau_9 \sim \text{Normal}(\rho q_{t(m-1)}, \tau_9^{-1})$, where $\rho$ is a mixing parameter and $\tau_9$ the precision. 
	
	\begin{table}[!t]
		\caption{The different specifications of the long-term temporal trend $\gamma_*$ parameter in the expected mortality model.}
		\label{tab1}
		\centering
		\begin{tabular}{lcl}
			\hline
			Models & Formulation of $\gamma_*$ & Note\\ 
			\hline
			1 & 0 & No long-term trend \\
			2 & $\beta_3 l$ & Linear long-term trend \\
			3 & $\beta_3 l + \beta_4 l I(l>c)$ & linear threshold on 2020-06-01\\
			4 & $f(l)$ & Natural cubic spline with 5 degrees of freedom\\
			5 & $\theta_t$ & $\theta_t \sim \text{Normal}(0, \tau_{11}^{-1})$ (IID)\\
			6 & $\eta_l$ &  $\eta_l \sim \text{Normal}(\rho_1\eta_{l-1} + \rho_2\eta_{l-2}, \tau_{12}^{-1})$ (AR2)\\
			7 & $\beta_3 l + \theta_t$ & Linear daily long trend and yearly IID\\
			8 & $\beta_3 l + \eta_l$ & Linear daily long trend and daily AR2\\
			\hline
		\end{tabular}
		\begin{flushleft} IID: Independent and Identically normally distributed, AR2: Autoregressive process of second order \end{flushleft}
		
	\end{table}
	
	To complete the Bayesian representation of our models we needed to specify priors. We considered improper priors for $\alpha_0$ and $\beta_0$, i.e. $\text{Normal}(0, \infty)$ and vague priors for the regression coefficients ($\alpha_1$ and $\beta_1, \dots, \beta_8$), i.e. $\text{Normal}(0, 1000)$. For the hyperparameters, we considered priors that penalise complexity from the null model \cite{simpsonpenal}. Specifically, we tuned the prior for the precision hyperparameters so that $\text{Pr}(1/\sqrt{\tau_p}<1) = 0.01, \; p = 1, \dots, 11$, the prior for the mixing term $\text{Pr}(\phi<0.5) = 0.5$ and the prior of the correlation $\text{Pr}(\rho<0.5) = 0.5$. For the prior of the precision of the autoregressive process $\eta_t$, Table SI17, we were stricter to ensure identifiability, $\text{Pr}(\sqrt{\tau_{12}}<0.01) = 0.01$, whereas for the correlation parameters we defined vague PC priors: $\text{Pr}(\rho_1<0.9) = 0.9$ and $\text{Pr}(\rho_2<0.9) = 0.9$.
	
	We trained the eight models using the years 2011–2021 and predict daily summer mortality for $t'=2022$ by canton. Each of the eight models were fitted for different population samples, $\xi=1, \dots, 200$ and then pooled together. We then sampled from the posterior predictive distributions:
	
	$$\text{p}_n(O_{st'lm}|\pmb{Q}) = \int \text{p}_n(O_{st'lm}|\pmb{h}) \text{p}_n(\pmb{h}|\pmb{Q})d\pmb{h}, \;\; n=1, \dots, 10$$
	\noindent where \pmb{h} is a vector of the model parameters and \pmb{Q} the observed data. We summarised the samples from the above posterior based on weights $\kappa_n$. We considered two types of weights: equal weights across the eight models (model 9) and bias-based weights as defined in the cross validation section in the next paragraph (model 10). We repeated the procedure for the different age and sex $g$ groups.

	\subsection*{Cross-validation}
	We performed a leave the last year out cross-validation to assess the performance of our models. For both population and expected mortality model we left the year 2021 out and assessed how well the models predicted it by assessing the bias, mean squared error (MSE) and coverage probability. We used a transformation of the age- and sex-specific bias (averaged over the entire summer 2021) by the $n$-th model to define weights $\kappa_{gn}$ for the pooled samples. Let $U_{gn}$ be the bias of the $n$-th model in the $g$-th group, then we defined weights as: $\kappa_{gn} = 1/U_{gn}$, giving higher weights to the models with smaller bias \citep{kontis2017future}.
	
	\subsection*{Heatwave-attributable mortality}
	To additionally examine the validity of our predictions we compared our estimated with deaths attributable to heatwaves. We defined a heat event as the period when the cantonal population-weighted mean temperatures are higher than $\delta = 22^oC$ (90-th percentile) for three consecutive days. We also considered 3-day lag was based on the literature \citep{heatwaveslag, ragettli2023explorative}. Following similar notation as before:
	\begin{align*}
		O_{stl} &\sim \text{Poisson}\big(\mu_{stl}\Tilde{P}_{stl}\big) \\
		\log(\mu_{stl}) &= \beta_{0} + \beta_1X_{1sl} + \beta_2 X_{2sl} + f(r_{stl}) + \sum_{j=1}^6(\beta_{3j}X_{3j}) + \beta_4 t + \\ & \quad (\beta_{50} + \beta_{5t})\prod_{k = 0}^2 I[r_{st(l-k)}>\delta] + z_{s} + w_{tl}
	\end{align*}
	\noindent where $\beta_{0}$ is an intercept, $\beta_{3j}$ is the effect of different days of the week $X_{3j}$, $\beta_4$ is a long term effect across the different years $t$, $\beta_{50}$ an overall heat-event effect that is allowed to vary by year through $\beta_{5t}$, $I(\cdot)$ an indicator function and $w_{tl}$ captures the temporal trends. We defined $\beta_{5t} \sim \text{Normal}(0, \tau_{13}^{-1})$ and $w_{tl} \sim \text{Normal}\big(2w_{t(l-1)} + w_{t(l-2)}, \tau_{14}^{-1}\big)$. The specification of the rest of the components (including the terms $\beta_1$, $\beta_2$, $f(r_{stl})$ and $z_{s}$) are identical with the expected mortality model.  Subsequently, we estimated the number of deaths attributable to these heat periods \citep{mansournia2018population, konstantinoudis2022ambient}: $$A_{stl} = \bigg(1 - 1/\exp\big(\beta_{50} + \beta_{5(t=t'}\big)\bigg)\sum_{stl}O_{stl}\big(\prod_{k = 0}^2 I[r_{st(l-k)}>\delta]\big). $$

	\subsection*{COVID-19 deaths during heatwaves}
	We also examined effect of heatwaves on COVID-19 mortality.  We followed a similar approach as the heatwave-attributable analysis model, changing the outcome to being COVID-19 deaths during 2021 and 2022. We excluded the year 2020 as COVID-19 deaths in 2020 are subject to changes in definition and different policies regarding testing and reporting \cite{karanikolos2020comparable}, year as a factor and weighted temperature as linear. We also estimated number of COVID-19 deaths attributable to a heatwave event as described before.
	
	\subsection*{Sensitivity analysis}
	As a sensitivity analysis we provided the heatwave effects and deaths attributable to heatwaves for the heatwave-attributable mortality without accounting for ambient temperature, as it could capture some of the effects of the heatwaves and vice versa. We also fitted the model for predicting expected mortality in summer 2022 without accounting for ambient temperature and report the excess number of deaths. 
	
	All models were fitted using the Integrated Nested Laplace Approximation (INLA) using its R software interface \citep{rue2009approximate, blangiardo2013spatial}. Code for running the analysis is available online (\url{https://github.com/gkonstantinoudis/excess-summer-ch}).

	\section{Results}
	
	\subsection*{Cross-validation}
	
	The different models predicting population size had an overall good performance, Supporting Information Appendix Figures SI1-3. The ensemble model with uniform weights performed better and was thus selected, Supporting Information Appendix Figures SI1-3. 
	The different models predicting daily mortality in summer 2021 showed good performance overall, with variations across the different age and sex groups, Supporting Information Appendix Tables SI1-10. The ensemble model with bias-related weights showed consistently good performance across the different age and sex groups and thus selected for the analysis, Supporting Information Appendix Figure SI4.
	
	\subsection*{Overall effect by age and sex}
	We observed in total 17,602 deaths during the summer months of 2022, Table \ref{tab2}. We estimated in total 513 (95\% Credible Interval:-48, 1039) excess deaths during the summer months of 2022, after accounting for national holidays, ambient temperature, COVID-19 deaths, population change and spatiotemporal trends in mortality. This is equivalent to 3\% (0\%, 6\%) relative excess mortality. The age group most affected was the oldest, with a total of 456 (5, 891) excess deaths during summer 2022. There is weak evidence to support excess mortality in the other age groups, Table \ref{tab2}. 
	
	\begin{table*}[!t]
		\caption{Expected, excess, relative excess and COVID-19 mortality during summer 2022 in Switzerland by age group and sex. For the estimated of expected, excess and relative excess mortality we show median and 95\% credible intervals.}
		\label{tab2}
		\centering
		\begin{tabular}{llrrrrr}
			\hline
			Age & Sex & Deaths & Expected & Excess & Relative Excess & COVID-19 \\ 
			\hline
			0-39 & female & 126 & 118 (93, 157) & 8 (-31, 33) & 0.07 (-0.26, 0.28) & 0 \\ 
			0-39 & male & 235 & 222 (193, 257) & 13 (-22, 42) & 0.06 (-0.10, 0.19) & 0 \\ 
			40-59 & female & 444 & 437 (383, 512) & 7 (-68, 61) & 0.02 (-0.16, 0.14) & 7 \\ 
			40-59 & male & 709 & 703 (643, 781) & 6 (-72, 66) & 0.01 (-0.10, 0.09) & 1 \\ 
			60-69 & female & 657 & 626 (581, 691) & 31 (-34, 76) & 0.05 (-0.05, 0.12) & 7 \\ 
			60-69 & male & 1070 & 1059 (950, 1187) & 11 (-117, 120) & 0.01 (-0.11, 0.11) & 16 \\ 
			70-79 & female & 1460 & 1464 (1364, 1580) & -4 (-120, 96) & 0.00 (-0.08, 0.07) & 24 \\ 
			70-79 & male & 2047 & 2033 (1883, 2237) & 14 (-190, 164) & 0.01 (-0.09, 0.08) & 44 \\ 
			80+ & female & 6326 & 6078 (5747, 6410) & 248 (-84, 579) & 0.04 (-0.01, 0.10) & 97 \\ 
			80+ & male & 4528 & 4307 (4083, 4619) & 221 (-91, 445) & 0.05 (-0.02, 0.10) & 89 \\ 
			0-39 & Total & 361 & 341 (300, 388) & 20 (-27, 61) & 0.06 (-0.08, 0.18) & 0 \\ 
			40-59 & Total & 1153 & 1148 (1060, 1251) & 5 (-98, 93) & 0.00 (-0.09, 0.08) & 8 \\ 
			60-69 & Total & 1727 & 1681 (1573, 1829) & 46 (-102, 154) & 0.03 (-0.06, 0.09) & 23 \\ 
			70-79 & Total & 3507 & 3500 (3317, 3800) & 7 (-293, 190) & 0.00 (-0.08, 0.05) & 68 \\ 
			80+ & Total & 10854 & 10398 (9963, 10849) & 456 (5, 891) & 0.04 ( 0.00, 0.09) & 186 \\ 
			Total & female & 9013 & 8729 (8361, 9104) & 284 (-91, 652) & 0.03 (-0.01, 0.07) & 135 \\ 
			Total & male & 8589 & 8362 (8010, 8756) & 227 (-167, 579) & 0.03 (-0.02, 0.07) & 150 \\ 
			Total & Total & 17602 & 17089 (16563, 17650) & 513 (-48, 1039) & 0.03 (0.00, 0.06) & 285 \\ 
			\hline
		\end{tabular}
	\end{table*}

	\subsection*{Temporal effects}
\begin{figure*}[!t]
		
		\centering
		\includegraphics[scale=0.9]{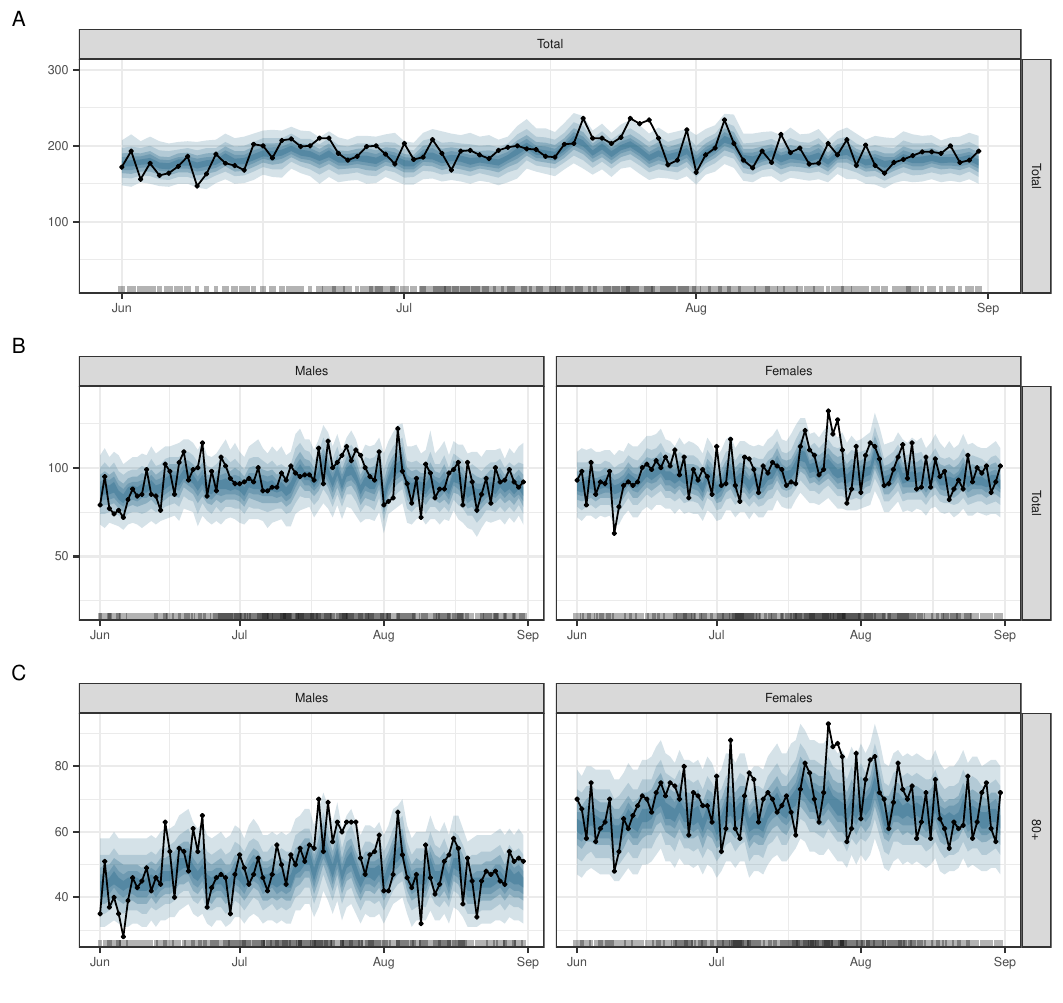}
		\caption{Daily observed and expected all-cause deaths overall (panel A), by sex (panel B) and by sex in age group 80$+$ (panel C) during summer 2022 in Switzerland. The shaded blue ribbon shows the 95\% credible interval of expected deaths. The black line shows the observed number of deaths. The black rug on the bottom of each plot denotes the number of COVID-19 deaths.}
		\label{fig1}
	\end{figure*}

	Figure \ref{fig1}A shows the daily observed and expected mortality during summer 2022. Overall the total mortality is in line with the expected with some excess observed after mid July. These observed patterns are similar for males and females and driven by people older than 80 years, Figure \ref{fig1}B. When we aggregated mortality by month, we reported 68 (-154, 276) excess deaths in June, 341 (113, 561) in July and 106 (-141, 335) in August for people aged 80 and older, Supporting Information Appendix Tables SI11-13.
	
	\subsection*{Spatial effects}
\begin{figure*}[!t]
		\centering
		\includegraphics[scale=0.9]{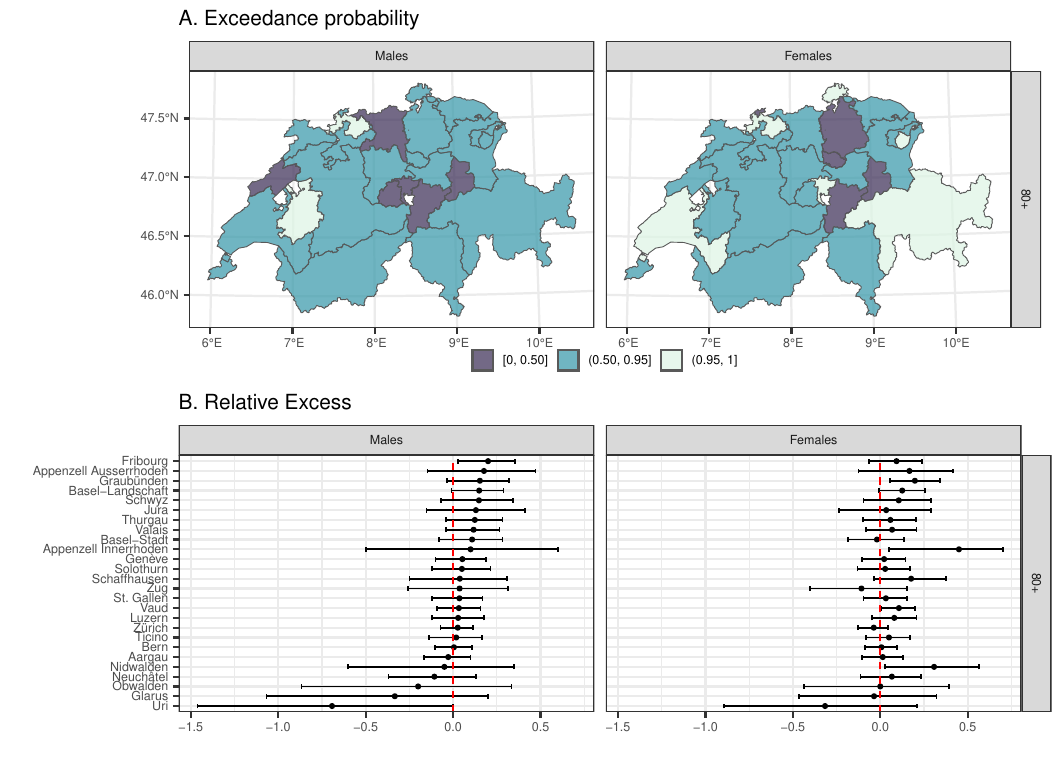}
		\caption{(A) Probability that the relative excess mortality in summer 2022 is higher than 0\% in Switzerland in populations older than 80 years old by canton. (B) Relative excess mortality  95\% credible interval by canton and sex across the summer 2022 in Switzerland for populations older than 80 years old.}
		\label{fig2}
	\end{figure*}

	The spatial distribution of excess mortality differed according to the sex, Figure \ref{fig2}. Among males, we found the highest evidence for non-null excess mortality in two cantons of the west (Fribourg) and the northwest (Basel-Landschaft). In females, there was also strong evidence for non-null excess in the same northwestern canton, but also in six other cantons scatterred around the country, Supporting Information Appendix Figure SI5. 
	
	\subsection*{Heatwave-attributable mortality}
	To further validate our estimates, we defined heatwave periods and compared the number of excess deaths occurring during heatwaves with the number of deaths attributable to heatwaves, regressing all-cause deaths with heatwave periods. We defined heatwave periods as 3 consecutive days of temperatures higher than the 90th percentile of the nationwide temperature with a three-day lagged effect to account for a delayed effect of heatwaves on mortality \cite{heatwaveslag, ragettli2023explorative}. We found that the estimated number of deaths attributable to heatwaves was consistent with our estimates of excess deaths during heatwaves, regardless of whether ambient temperature was considered, Supporting Information Appendix Table SI14. We observed a decreasing trend over time across the different age groups, suggesting progressive adaptation of the population and highlighting the importance for accounting for temporal variation in the effect of temperature on mortality, Supporting Information Appendix Figures SI6-7.
	
	We considered 95 alternative definitions for a heatwave, with temperature thresholds varying between the 5th and the 95th percentile of the nationwide temperature for a duration varying between 1 and 5 consecutive days, Figure \ref{fig3} and Supporting Information Appendix Figure SI8. For each combination, we computed the number of excess deaths and the relative excess mortality during summer 2022. We then calculated minimum excess mortality thresholds, defined as the temperature at which the probability of non-null excess mortality is higher than 0.95. We estimated this minimum excess mortality threshold to 19$^o$C (70th percentile) for heatwaves defined over four or five days, and 21$^o$C (85th percentile) for heatwaves defined over three days, Figure \ref{fig3}). We found limited evidence of an increase in mortality for heatwaves lasting 1 or 2 days, reflecting the adjustment for ambient temperature in the excess mortality model but also the fact that exposure was higher compared to historical levels in 2022.
	
	\begin{figure*}[!t]
		\centering
		\includegraphics{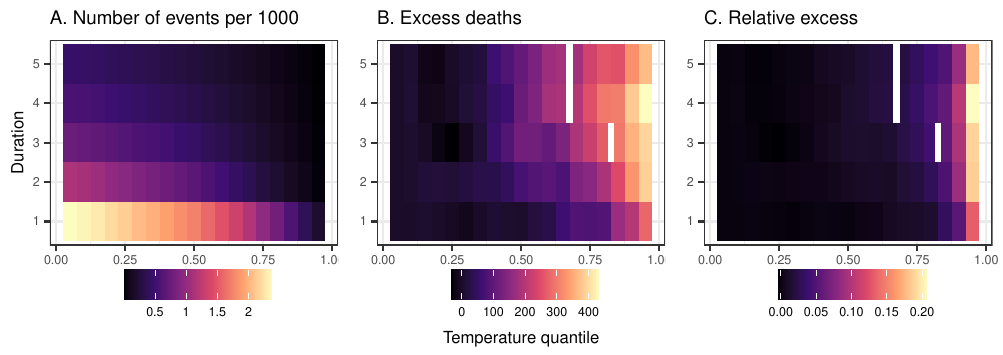}
		\caption{(A) Number of heatwave event per 1,000 based on 95 alternative definitions based on thresholds of temperature (in percentile of the nationwide temperature, x-axis) and duration (y-axis). (B-C). Number of excess deaths and relative excess mortality across the different heatwave definitions. The thick white segment corresponds to the minimum excess temperature mortality threshold, i.e. the temperature at which the probability of non-null excess mortality is higher than 0.95.}
		\label{fig3}
	\end{figure*}

	\subsection*{COVID-19 deaths during heatwaves}
	We investigated the existence of an interaction between COVID-19 and extreme heat events by comparing COVID-19 deaths occurring during and outside of heatwaves (using the initial definition of heatwaves). We observed a 1.43 (0.84, 2.45) and 1.84 (1.16, 2.90) increase in the COVID-19 mortality rate in males and females older than 80 years respectively, Supporting Information Appendix Table SI15. Conditional on our definition of a heatwave and assuming causality, this leads to 15 (5, 22) deaths attributable to heatwaves in females over 80 years old.
	
	\subsection*{Sensitivity analysis}
	Not accounting for ambient temperature led to, as expected, higher excess deaths during summer 2022, namely 628 (23, 1162) in total, Supporting Information Appendix Table SI16. This number includes deaths due to non-extreme heat, which happen every year. The main goal of our analysis is to capture deaths due to the extreme and prolonged heat events during summer 2022.

	\section{Discussion}

	In this nationwide study in Switzerland and after accounting for national holidays, ambient temperature, COVID-19 deaths, population change and other spatiotemporal trends, we observed a 3\% (0\%, 6\%) relative excess mortality during summer 2022 in Switzerland. This number is almost two times lower than estimated during summer 2003 in Switzerland \citep{thommen2005heat}. The evidence was strongest in people 80 years or older and the effect was similar across males and females. We observed weak evidence of excess mortality in the other age groups. When examining the temporal trends in excess mortality, July was the month with the highest excess mortality. We observed different spatial trends in excess mortality across the different sexes. We lastly calculated the minimum excess mortality temperature threshold and reported that for heatwaves with longer duration (4 or 5 days), this threshold is the 70th percentile of the nationwide temperature (19$^o$C).  
	
	Our results are comparable with three previous studies in Switzerland estimating mortality attributable to heat stress\citep{vicedo2023footprint, raglettiMonitoring2023, Ballester2023}. The first study estimated heat-related mortality attributable to human-induced warming as the difference between the deaths due to the observed temperatures and temperatures based on the counterfactual scenario of a climate in absence of an anthropogenic signal between June and August 2022 \cite{vicedo2023footprint}. It had high spatial resolution for both outcome (residential location) and exposure (temperature by 2 km-cell) and a long study period (1990-2017). The authors  reported 623 (151, 1068) deaths due to heat between June and August 2022, with almost 90\% of the estimated heat-related mortality in older adults \cite{vicedo2023footprint}. The second study had coarser exposure resolution, i.e., temperature measurements per meteorological station in Switzerland (7 stations, one for each large region), focused on May until September 2022 and reported 474 (271, 674) deaths attributed to heat stress, with most deaths in populations older than 75 years \cite{raglettiMonitoring2023}. The third study used weekly counts of all-cause mortality according to sex, age and cantons and temperature exposure at 9km grid \cite{Ballester2023}. They reported 302 (48, 557) deaths attributable to heat in Switzerland during June-August 2023 \cite{Ballester2023}. The results of these studies depend on the granularity of the resolution of the outcome and exposure, with higher resolution resulting in a higher mortality burden \cite{vicedo2023footprint, raglettiMonitoring2023, Ballester2023}. In our study, a more granular spatiotemporal resolution would only affect the flexibility to aggregate the results in different spatiotemporal scales and it is not expected to affect the number of excess deaths. We report 513 (-48, 1039) excess deaths in summer 2022, with the older populations being affected the most. The credible intervals of our estimates are large and compatible with all previous studies, reflecting the complex temporal trends in the mortality accounted using the ensemble model and the population uncertainty.
	
	Our approach has several strengths. It is a Bayesian ensemble method, consisting of multiple models with different assumptions regarding the shape of temporal trends across age groups, sex and space. This can make it an automated tool for mortality surveillance, that does not rely on a single "true" model. In addition to that, our approach is driven by the observed levels of mortality during the period of interest, and does not make any assumptions regarding causal relationships, adaptation and vulnerable subgroups. We proposed a method for calculating dynamic population, accounting for population changes because of deaths occurring during the extreme events themselves. Not accounting for this will likely increase the estimates of excess deaths, as the population counts will be larger. Our approach is invariant to the different age, sex and spatiotemporal scales, as results can be aggregated at any level, providing information about different vulnerabilities. As there is no unified approach for defining a heatwave, results can be also aggregated in ways to inform policies about relevant thresholds and periods of heat alerts. This is the first study in Switzerland that accounts for the effect of COVID-19 when estimating the excess deaths during summer 2022, a period marked by epidemics waves linked to the emergence of variants omicron BA-4 and BA-5.
	
	Our study has some limitations. We accounted for the effect of COVID-19 in our estimates by estimating its contribution to all cause deaths during summers 2020-21. This assumes that COVID-19 deaths ascertainment is complete in these periods, which might be a strong assumption especially for the early stages of the pandemic \citep{karanikolos2020comparable}. We modelled the population dynamically using information on daily deaths, however we did not have information on daily migration and birth, which were assumed to change with a constant rate across the year. In line with other studies of excess mortality to quantify the effect of COVID-19 pandemic \citep{konstantinoudis2022regional, riou2023direct, wang2022estimating}, we cannot definitely attribute the excess mortality on a specific source, although we can rule out the short-term effect of COVID-19 as we accounted for COVID-19 deaths in our estimates.
	
	Using the developed modelling framework, we can not attribute deaths directly to exposure to prolonged periods of heat. Other potential candidates for the observed excess include the effect of COVID-19. This could translate to a short- (COVID-19 deaths) or long-term effect (lagged effect of the COVID-19 pandemic, for instance due to delays in cancer screening during the early stages \cite{riera2021delays}). In our analysis, we have accounted for the short-term effect of COVID-19 deaths but cannot rule out residual confounding due to inadequate adjustment. We did not account for any lagged effect of the COVID-19 pandemic, which could have contributed to an overall increased mortality during summer 2022, but probably not period-specific. Another potential candidate is the aftermath of the war in Ukraine, but this event is unlikely to be linked with higher mortality in Switzerland. Based on the facts that (i) we accounted for the short-term effect of COVID-19, (ii) the observed excess mortality was highest in the older populations, and (iii) that the excess mortality is in line with periods of extensive Figure SI18-19, Tables SI11-13 and Figure \ref{fig3}. For these reasons, we argue that the main candidate of the observed excess mortality are the extreme heat events of summer 2022.
	
	Our study also provides information about relevant thresholds for heat warnings. In 2021, MeteoSwiss revised the heat warning system issuing a level 2 warning as soon as the mean daily temperature is higher than 25$^o$C, a level 3 if it is higher than 25$^o$C for at least three consecutive days and a level 4 it is higher than 27$^o$C for at least three consecutive days \citep{meteoswiss2021}. These thresholds are based on a paper using high resolution nationwide mortality (residential address) and temperature data (100m) \citep{ragettli2023explorative}. In our study we confirm the importance of the three consecutive days, but find a minimum mortality temperature threshold at around $20^o$C \citep{ragettli2023explorative}. Nevertheless, this minimum excess mortality temperature threshold decreases as the duration of the heatwave is longer. As our evidence of excess mortality was only strong for populations over 80 years old, we consider these results relevant for this age subgroup. We stress that our recommendations are based on the cantonal population weighted mean temperatures using mean temperatures at 2m from the ERA-5 reanalysis dataset. Future studies should focus on warning regions and the relevant meteorological stations considered for the heat warnings and evaluate the validity of these thresholds in light of the policy implemented. 
	
	In conclusion, this study provides a comprehensive analysis of daily excess mortality during the summer 2022 in Switzerland disaggregated by sex and age groups and cantons. We found increased COVID-19 mortality during heatwaves, suggesting that individuals with COVID-19 might be a vulnerable group. We have proposed a flexible model, thoroughly validated, that can be used as an automated tool for monitoring extreme temperature events and informing about thresholds relevant for policy-makers. Given the newly revised definition of heat warnings and after accounting for COVID-19, we observed increased excess deaths during summer 2022, mainly in people older than 80 years old. To reduce future summer excess mortality in Switzerland, we propose targeted heat warnings to older populations and reducing the temperature threshold when forecasts suggest periods of extreme heat longer than 4 days.

	\section*{Data availability}
	We provide at \url{https://github.com/gkonstantinoudis/excess-summer-ch} the final version of the datasets.
	
	Raw mortality data on all-cause and COVID-19 deaths can be provided upon request from the Federal Office of Statistics in Switzerland \url{https://www.bfs.admin.ch/bfs/en/home.html}. Raw population files can be found here: \url{https://www.pxweb.bfs.admin.ch/pxweb/en/px-x-0102010000_102/-/px-x-0102010000_102.px/}. Air temperature at 2m for was retrieved from \url{https://cds.climate.copernicus.eu/cdsapp#!/dataset/reanalysis-era5-land?tab=form}.

	\section*{Code and Reproducibility}
	All models were fitted using the Integrated Nested Laplace Approximation (INLA) using its R software interface \cite{rue2009approximate}.
	To ensure reproducibility and transparency to our results and approach the code for running the analysis is available at \url{https://github.com/gkonstantinoudis/excess-summer-ch}. 
	
	\section*{Ethics}
	The study is about secondary, aggregate anonymised data so no ethical permission is required.
	
	\section*{Acknowledgements}
	We would like to thank Cordula Blohm and the Federal Statistical Office in Switzerland for curating and sharing the mortality data.
	
	G.K. is supported by an MRC Skills Development Fellowship [MR/T025352/1] and an Imperial College Research Fellowship. J.R. is supported by the Swiss Federal Office of Public Health (142005806) and by the Swiss National Science Foundation (189498).
	
	\section*{Author contributions}
	G.K. conceived the study. J.R. and G.K. supervised the study. G.K. developed the initial study protocol and discussed it with J.R. and A.H.. G.K. developed the statistical model, prepared the population and covariate data and led the acquisition of all cause mortality data. J.R. led the acquisition of COVID-19 mortality data. A.H. calculated the mean temperature weighted by population. A.H., J.R. and G.K. conceived the dynamic population adjustment. G.K. ran the analysis and wrote the initial draft. All the authors contributed in modifying the paper and critically interpreting the results. All authors read and approved the final version for publication.
	
	\section*{Competing interests}
	The authors declare no competing interests.

	\bibliographystyle{unsrt}
	\bibliography{bib.bib}
	
	\FloatBarrier
	
	\clearpage
	
	\setcounter{table}{0}
	\setcounter{figure}{0}
	
	\renewcommand{\thetable}{SI\arabic{table}}  
	\renewcommand{\thefigure}{SI\arabic{figure}}
	
	\renewcommand{\figurename}{Figure}
	\renewcommand{\tablename}{Table}
	
	\section*{Supporting Information Appendix}
	
	\begin{table}[ht]
		\caption{Mean square error (MSE), bias and coverage probability for \textbf{males} aged between \textbf{0 and 39 years old} across the 10 models for estimating excess mortality.}
		\centering
		\begin{tabular}{llllll}
			\hline
			Age & Sex & Model & MSE & Bias & Coverage \\ 
			\hline
			0-39 & Males & Model 1 & 0.19 (0.17, 0.21) &  0.00 (-0.01,  0.02) & 0.67 \\ 
			0-39 & Males & Model 2 & 0.17 (0.16, 0.20) & -0.01 (-0.02,  0.01) & 0.65 \\ 
			0-39 & Males & Model 3 & 0.18 (0.15, 0.23) &  0.00 (-0.03,  0.04) & 0.67 \\ 
			0-39 & Males & Model 4 & 0.20 (0.16, 0.34) &  0.02 (-0.02,  0.12) & 0.72 \\ 
			0-39 & Males & Model 5 & 0.19 (0.17, 0.21) &  0.00 (-0.01,  0.02) & 0.68 \\ 
			0-39 & Males & Model 6 & 0.19 (0.17, 0.21) &  0.00 (-0.01,  0.02) & 0.68 \\ 
			0-39 & Males & Model 7 & 0.18 (0.16, 0.20) & -0.01 (-0.02,  0.01) & 0.65 \\ 
			0-39 & Males & Model 8 & 0.18 (0.16, 0.20) & -0.01 (-0.02,  0.01) & 0.65 \\ 
			0-39 & Males & Model 9 & 0.18 (0.16, 0.22) &  0.00 (-0.02,  0.04) & 0.68 \\ 
			0-39 & Males & Model 10 & 0.18 (0.16, 0.25) &  0.00 (-0.02,  0.06) & 0.67 \\ 
			\hline
		\end{tabular}
	\end{table}
	MSE: Mean Square Error
	\clearpage
	
	\begin{table}[ht]
		\caption{Mean square error (MSE), bias and coverage probability for \textbf{males} aged between \textbf{40 and 59 years old} across the 10 models for estimating excess mortality.}
		\centering
		\begin{tabular}{llllll}
			\hline
			age & sex & Model & MSE & Bias & Coverage \\ 
			\hline
			40-59 & Males & Model 1 & 0.62 (0.58, 0.69) &  0.02 ( 0.00,  0.04) & 0.88 \\ 
			40-59 & Males & Model 2 & 0.59 (0.54, 0.64) & -0.01 (-0.04,  0.02) & 0.85 \\ 
			40-59 & Males & Model 3 & 0.60 (0.53, 0.71) &  0.00 (-0.06,  0.07) & 0.86 \\ 
			40-59 & Males & Model 4 & 0.58 (0.50, 0.68) & -0.03 (-0.09,  0.05) & 0.84 \\ 
			40-59 & Males & Model 5 & 0.62 (0.56, 0.71) &  0.02 (-0.02,  0.07) & 0.87 \\ 
			40-59 & Males & Model 6 & 0.62 (0.57, 0.68) &  0.02 ( 0.00,  0.05) & 0.88 \\ 
			40-59 & Males & Model 7 & 0.58 (0.54, 0.64) & -0.01 (-0.04,  0.01) & 0.85 \\ 
			40-59 & Males & Model 8 & 0.59 (0.54, 0.64) & -0.01 (-0.04,  0.02) & 0.86 \\ 
			40-59 & Males & Model 9 & 0.60 (0.52, 0.69) & -0.01 (-0.06,  0.05) & 0.85 \\ 
			40-59 & Males & Model 10 & 0.60 (0.53, 0.67) &  0.00 (-0.06,  0.04) & 0.86 \\ 
			\hline
		\end{tabular}
	\end{table}
	MSE: Mean Square Error
	
	\clearpage
	
	\begin{table}[ht]
		\caption{Mean square error (MSE), bias and coverage probability for \textbf{males} aged between \textbf{60 and 69 years old} across the 10 models for estimating excess mortality.}
		\centering
		\begin{tabular}{llllll}
			\hline
			age & sex & Model & MSE & Bias & Coverage \\ 
			\hline
			60-69 & Males & Model 1 & 0.89 (0.83, 0.98) &  0.04 ( 0.01,  0.07) & 0.92 \\ 
			60-69 & Males & Model 2 & 0.82 (0.77, 0.88) & -0.03 (-0.06,  0.00) & 0.90 \\ 
			60-69 & Males & Model 3 & 0.79 (0.73, 0.87) & -0.08 (-0.14,  0.00) & 0.88 \\ 
			60-69 & Males & Model 4 & 0.78 (0.71, 0.87) & -0.10 (-0.17, -0.01) & 0.86 \\ 
			60-69 & Males & Model 5 & 0.89 (0.79, 1.05) &  0.04 (-0.05,  0.15) & 0.92 \\ 
			60-69 & Males & Model 6 & 0.89 (0.83, 0.97) &  0.04 ( 0.01,  0.07) & 0.92 \\ 
			60-69 & Males & Model 7 & 0.82 (0.76, 0.89) & -0.03 (-0.06,  0.00) & 0.90 \\ 
			60-69 & Males & Model 8 & 0.82 (0.76, 0.89) & -0.03 (-0.07,  0.01) & 0.89 \\ 
			60-69 & Males & Model 9 & 0.83 (0.76, 0.96) & -0.02 (-0.08,  0.07) & 0.90 \\ 
			60-69 & Males & Model 10 & 0.83 (0.74, 0.96) & -0.03 (-0.15,  0.07) & 0.90 \\ 
			\hline
		\end{tabular}
	\end{table}
	MSE: Mean Square Error
	
	\clearpage
	
	\begin{table}[ht]
		\caption{Mean square error (MSE), bias and coverage probability for \textbf{males} aged between \textbf{70 and 79 years old} across the 10 models for estimating excess mortality.}
		\centering
		\begin{tabular}{llllll}
			\hline
			age & sex & Model & MSE & Bias & Coverage \\ 
			\hline 
			70-79 & Males & Model 1 & 1.67 (1.56, 1.84) &  0.05 ( 0.02,  0.09) & 0.93 \\ 
			70-79 & Males & Model 2 & 1.56 (1.46, 1.65) & -0.06 (-0.11, -0.02) & 0.92 \\ 
			70-79 & Males & Model 3 & 1.59 (1.47, 1.75) & -0.02 (-0.12,  0.07) & 0.92 \\ 
			70-79 & Males & Model 4 & 1.51 (1.42, 1.65) & -0.11 (-0.24, -0.01) & 0.90 \\ 
			70-79 & Males & Model 5 & 1.68 (1.47, 1.96) &  0.04 (-0.10,  0.21) & 0.93 \\ 
			70-79 & Males & Model 6 & 1.68 (1.56, 1.81) &  0.05 ( 0.01,  0.09) & 0.93 \\ 
			70-79 & Males & Model 7 & 1.55 (1.45, 1.69) & -0.06 (-0.11, -0.02) & 0.92 \\ 
			70-79 & Males & Model 8 & 1.56 (1.47, 1.68) & -0.06 (-0.13,  0.01) & 0.91 \\ 
			70-79 & Males & Model 9 & 1.61 (1.48, 1.79) & -0.02 (-0.13,  0.14) & 0.92 \\ 
			70-79 & Males & Model 10 & 1.58 (1.46, 1.77) & -0.04 (-0.16,  0.14) & 0.93 \\ 
			\hline
		\end{tabular}
	\end{table}
	MSE: Mean Square Error
	
	\clearpage
	
	\begin{table}[ht]
		\caption{Mean square error (MSE), bias and coverage probability for \textbf{males} aged \textbf{over 80 years old} across the 10 models for estimating excess mortality.}
		\centering
		\begin{tabular}{llllll}
			\hline
			age & sex & Model & MSE & Bias & Coverage \\ 
			\hline
			80+ & Males & Model 1 & 3.59 (3.33, 3.87) &  0.15 ( 0.09,  0.21) & 0.94 \\ 
			80+ & Males & Model 2 & 3.39 (3.13, 3.67) &  0.00 (-0.07,  0.07) & 0.92 \\ 
			80+ & Males & Model 3 & 3.48 (3.18, 3.87) &  0.06 (-0.09,  0.22) & 0.94 \\ 
			80+ & Males & Model 4 & 3.43 (3.20, 3.81) &  0.02 (-0.16,  0.24) & 0.93 \\ 
			80+ & Males & Model 5 & 3.61 (3.17, 4.17) &  0.14 (-0.06,  0.34) & 0.94 \\ 
			80+ & Males & Model 6 & 3.60 (3.36, 3.87) &  0.15 ( 0.09,  0.20) & 0.94 \\ 
			80+ & Males & Model 7 & 3.41 (3.18, 3.63) &  0.00 (-0.06,  0.07) & 0.93 \\ 
			80+ & Males & Model 8 & 3.40 (3.15, 3.70) &  0.00 (-0.15,  0.11) & 0.93 \\ 
			80+ & Males & Model 9 & 3.40 (3.14, 3.69) &  0.01 (-0.08,  0.10) & 0.92 \\ 
			80+ & Males & Model 10 & 3.45 (3.17, 3.83) &  0.04 (-0.08,  0.24) & 0.93 \\ 
			\hline
		\end{tabular}
	\end{table}
	MSE: Mean Square Error
	
	\clearpage
	
	\begin{table}[ht]
		\caption{Mean square error (MSE), bias and coverage probability for \textbf{females} aged between \textbf{0 and 39 years old} across the 10 models for estimating excess mortality.}
		\centering
		\begin{tabular}{llllll}
			\hline
			age & sex & Model & MSE & Bias & Coverage \\ 
			\hline
			0-39 & Females & Model 1 & 0.09 (0.08, 0.10) &  0.01 ( 0.00,  0.02) & 0.56 \\ 
			0-39 & Females & Model 2 & 0.09 (0.07, 0.10) &  0.00 (-0.01,  0.01) & 0.54 \\ 
			0-39 & Females & Model 3 & 0.08 (0.06, 0.11) &  0.00 (-0.02,  0.03) & 0.50 \\ 
			0-39 & Females & Model 4 & 0.08 (0.06, 0.14) &  0.00 (-0.02,  0.05) & 0.54 \\ 
			0-39 & Females & Model 5 & 0.09 (0.08, 0.10) &  0.01 (-0.01,  0.02) & 0.55 \\ 
			0-39 & Females & Model 6 & 0.09 (0.08, 0.10) &  0.01 ( 0.00,  0.02) & 0.55 \\ 
			0-39 & Females & Model 7 & 0.08 (0.07, 0.10) &  0.00 (-0.01,  0.01) & 0.53 \\ 
			0-39 & Females & Model 8 & 0.09 (0.07, 0.10) &  0.00 (-0.01,  0.02) & 0.53 \\ 
			0-39 & Females & Model 9 & 0.08 (0.06, 0.12) &  0.00 (-0.02,  0.03) & 0.52 \\ 
			0-39 & Females & Model 10 & 0.08 (0.07, 0.12) &  0.00 (-0.02,  0.03) & 0.53 \\ 
			\hline
		\end{tabular}
	\end{table}
	MSE: Mean Square Error
	
	\clearpage
	
	\begin{table}[ht]
		\caption{Mean square error (MSE), bias and coverage probability for \textbf{females} aged between \textbf{40 and 59 years old} across the 10 models for estimating excess mortality.}
		\centering
		\begin{tabular}{llllll}
			\hline
			age & sex & Model & MSE & Bias & Coverage \\ 
			\hline
			40-59 & Females & Model 1 & 0.36 (0.33, 0.43) &  0.01 (-0.01,  0.03) & 0.79 \\ 
			40-59 & Females & Model 2 & 0.34 (0.30, 0.42) & -0.02 (-0.04,  0.00) & 0.74 \\ 
			40-59 & Females & Model 3 & 0.33 (0.29, 0.40) & -0.03 (-0.07,  0.01) & 0.72 \\ 
			40-59 & Females & Model 4 & 0.33 (0.28, 0.43) & -0.03 (-0.08,  0.03) & 0.72 \\ 
			40-59 & Females & Model 5 & 0.37 (0.31, 0.46) &  0.01 (-0.04,  0.06) & 0.78 \\ 
			40-59 & Females & Model 6 & 0.37 (0.33, 0.44) &  0.01 (-0.01,  0.03) & 0.78 \\ 
			40-59 & Females & Model 7 & 0.33 (0.31, 0.41) & -0.02 (-0.04,  0.00) & 0.74 \\ 
			40-59 & Females & Model 8 & 0.34 (0.30, 0.38) & -0.02 (-0.04,  0.01) & 0.75 \\ 
			40-59 & Females & Model 9 & 0.36 (0.31, 0.44) &  0.00 (-0.04,  0.04) & 0.78 \\ 
			40-59 & Females & Model 10 & 0.34 (0.29, 0.42) & -0.02 (-0.07,  0.03) & 0.75 \\ 
			\hline
		\end{tabular}
	\end{table}
	MSE: Mean Square Error
	
	\clearpage
	
	\begin{table}[ht]
		\caption{Mean square error (MSE), bias and coverage probability for \textbf{females} aged between \textbf{60 and 69 years old} across the 10 models for estimating excess mortality.}
		\centering
		\begin{tabular}{llllll}
			\hline
			age & sex & Model & MSE & Bias & Coverage \\ 
			\hline
			60-69 & Females & Model 1 & 0.58 (0.52, 0.80) &  0.03 ( 0.01,  0.05) & 0.88 \\ 
			60-69 & Females & Model 2 & 0.54 (0.48, 0.74) &  0.00 (-0.03,  0.03) & 0.84 \\ 
			60-69 & Females & Model 3 & 0.53 (0.45, 0.80) & -0.02 (-0.06,  0.05) & 0.84 \\ 
			60-69 & Females & Model 4 & 0.50 (0.42, 0.73) & -0.05 (-0.10,  0.04) & 0.80 \\ 
			60-69 & Females & Model 5 & 0.58 (0.49, 0.91) &  0.03 (-0.02,  0.08) & 0.87 \\ 
			60-69 & Females & Model 6 & 0.58 (0.51, 0.81) &  0.03 ( 0.01,  0.06) & 0.87 \\ 
			60-69 & Females & Model 7 & 0.54 (0.48, 0.85) &  0.00 (-0.02,  0.03) & 0.85 \\ 
			60-69 & Females & Model 8 & 0.54 (0.47, 0.85) &  0.00 (-0.03,  0.03) & 0.85 \\ 
			60-69 & Females & Model 9 & 0.54 (0.47, 0.79) &  0.00 (-0.03,  0.03) & 0.84 \\ 
			60-69 & Females & Model 10 & 0.56 (0.46, 0.79) &  0.00 (-0.08,  0.05) & 0.85 \\ 
			\hline
		\end{tabular}
	\end{table}
	MSE: Mean Square Error
	
	\clearpage
	
	\begin{table}[ht]
		\caption{Mean square error (MSE), bias and coverage probability for \textbf{females} aged between \textbf{70 and 79 years old} across the 10 models for estimating excess mortality.}
		\centering
		\begin{tabular}{llllll}
			\hline
			age & sex & Model & MSE & Bias & Coverage \\ 
			\hline
			70-79 & Females & Model 1 & 1.17 (1.09, 1.28) &  0.00 (-0.03,  0.04) & 0.92 \\ 
			70-79 & Females & Model 2 & 1.13 (1.04, 1.24) & -0.03 (-0.07,  0.01) & 0.90 \\ 
			70-79 & Females & Model 3 & 1.12 (1.03, 1.24) & -0.06 (-0.13,  0.04) & 0.90 \\ 
			70-79 & Females & Model 4 & 1.08 (1.01, 1.20) & -0.14 (-0.21, -0.01) & 0.86 \\ 
			70-79 & Females & Model 5 & 1.17 (1.07, 1.29) &  0.00 (-0.07,  0.07) & 0.91 \\ 
			70-79 & Females & Model 6 & 1.17 (1.10, 1.26) &  0.00 (-0.03,  0.04) & 0.91 \\ 
			70-79 & Females & Model 7 & 1.14 (1.06, 1.22) & -0.03 (-0.07,  0.01) & 0.90 \\ 
			70-79 & Females & Model 8 & 1.14 (1.05, 1.24) & -0.03 (-0.09,  0.02) & 0.90 \\ 
			70-79 & Females & Model 9 & 1.17 (1.08, 1.29) &  0.00 (-0.05,  0.07) & 0.91 \\ 
			70-79 & Females & Model 10 & 1.14 (1.04, 1.27) & -0.03 (-0.17,  0.04) & 0.90 \\ 
			\hline
		\end{tabular}
	\end{table}
	MSE: Mean Square Error
	
	\clearpage
	
	\begin{table}[ht]
		\caption{Mean square error (MSE), bias and coverage probability for \textbf{females} aged \textbf{over 80 years old} across the 10 models for estimating excess mortality.}
		\centering
		\begin{tabular}{llllll}
			\hline
			age & sex & Model & MSE & Bias & Coverage \\ 
			\hline
			80+ & Females & Model 1 & 4.64 (4.26, 5.05) &  0.16 ( 0.09,  0.23) & 0.95 \\ 
			80+ & Females & Model 2 & 4.49 (4.15, 4.82) &  0.03 (-0.05,  0.12) & 0.94 \\ 
			80+ & Females & Model 3 & 4.46 (4.12, 4.84) & -0.02 (-0.19,  0.17) & 0.94 \\ 
			80+ & Females & Model 4 & 4.44 (4.14, 4.74) & -0.23 (-0.44,  0.03) & 0.92 \\ 
			80+ & Females & Model 5 & 4.64 (4.23, 5.21) &  0.15 (-0.04,  0.37) & 0.95 \\ 
			80+ & Females & Model 6 & 4.63 (4.31, 5.03) &  0.15 ( 0.09,  0.21) & 0.95 \\ 
			80+ & Females & Model 7 & 4.50 (4.24, 4.84) &  0.02 (-0.04,  0.10) & 0.95 \\ 
			80+ & Females & Model 8 & 4.51 (4.15, 4.92) &  0.04 (-0.07,  0.15) & 0.95 \\ 
			80+ & Females & Model 9 & 4.51 (4.20, 4.85) &  0.02 (-0.16,  0.16) & 0.94 \\ 
			80+ & Females & Model 10 & 4.55 (4.19, 4.91) &  0.04 (-0.31,  0.21) & 0.94 \\ 
			\hline
		\end{tabular}
	\end{table}
	MSE: Mean Square Error
	
	\clearpage
	
	\begin{table}[ht]
		\caption{Expected, excess, relative excess and COVID-19 mortality during \textbf{June} 2022 in Switzerland by age and sex. For the estimated of expected, excess and relative excess mortality we show median and 95\% credible intervals (The probability that the truth lies inside this interval is 0.95).}
		\centering
		\begin{tabular}{llrrrrrr}
			\hline
			Age & Sex & Deaths & Expected & Excess & Relative Excess & COVID-19 \\
			\hline
			0-39 & female & 30 & 37 (27, 49) & -7 (-19, 3) & -0.19 (-0.51, 0.08) & 0 \\ 
			0-39 & male & 76 & 72 (55, 88) & 4 (-12, 21) &  0.06 (-0.17, 0.29) & 0 \\ 
			40-59 & female & 136 & 144 (115, 173) & -8 (-37, 21) & -0.06 (-0.26, 0.15) & 2 \\ 
			40-59 & male & 229 & 228 (203, 259) & 1 (-30, 26) &  0.00 (-0.13, 0.11) & 0 \\ 
			60-69 & female & 188 & 206 (170, 236) & -18 (-48, 18) & -0.09 (-0.23, 0.09) & 1 \\ 
			60-69 & male & 360 & 343 (304, 393) & 17 (-33, 56) &  0.05 (-0.10, 0.16) & 2 \\ 
			70-79 & female & 462 & 469 (429, 519) & -7 (-57, 33) & -0.01 (-0.12, 0.07) & 2 \\ 
			70-79 & male & 680 & 663 (600, 745) & 17 (-65, 80) &  0.03 (-0.10, 0.12) & 12 \\ 
			80+ & female & 2006 & 1928 (1808, 2064) & 78 (-58, 198) &  0.04 (-0.03, 0.10) & 19 \\ 
			80+ & male & 1375 & 1373 (1277, 1489) & 2 (-114, 98) &  0.00 (-0.08, 0.07) & 14 \\ 
			0-39 & Total & 106 & 110 (87, 132) & -4 (-26, 19) & -0.04 (-0.24, 0.17) & 0 \\ 
			40-59 & Total & 365 & 373 (334, 412) & -8 (-47, 31) & -0.02 (-0.13, 0.08) & 2 \\ 
			60-69 & Total & 548 & 546 (496, 607) & 2 (-59, 52) &  0.00 (-0.11, 0.10) & 3 \\ 
			70-79 & Total & 1142 & 1134 (1053, 1250) & 8 (-108, 89) &  0.01 (-0.10, 0.08) & 14 \\ 
			80+ & Total & 3381 & 3305 (3144, 3481) & 76 (-100, 237) &  0.02 (-0.03, 0.07) & 33 \\ 
			Total & female & 2822 & 2784 (2645, 2938) & 38 (-116, 177) & 0.01 (-0.04, 0.06) & 24 \\ 
			Total & male & 2720 & 2687 (2545, 2841) & 33 (-121, 175) & 0.01 (-0.05, 0.07) & 28 \\ 
			Total & Total & 5542 & 5474 (5266, 5696) & 68 (-154, 276) &  0.01 (-0.03,  0.05) & 52 \\ 
			\hline
		\end{tabular}
	\end{table}
	
	\clearpage
	
	\begin{table}[ht]
		\caption{Expected, excess, relative excess and COVID-19 mortality during \textbf{July} 2022 in Switzerland by age and sex. For the estimated of expected, excess and relative excess mortality we show median and 95\% credible intervals (The probability that the truth lies inside this interval is 0.95).}
		\centering
		\begin{tabular}{llrrrrrr}
			\hline
			Age & Sex & Deaths & Expected & Excess & Relative Excess & COVID-19 \\ 
			\hline
			0-39 & female & 49 & 41 (30, 57) & 8 (-8, 19) &  0.20 (-0.20, 0.46) & 0 \\ 
			0-39 & male & 83 & 74 (61, 91) & 9 (-8, 22) &  0.12 (-0.11, 0.30) & 0 \\ 
			40-59 & female & 158 & 149 (124, 179) & 9 (-21, 34) &  0.06 (-0.14, 0.23) & 5 \\ 
			40-59 & male & 245 & 241 (206, 275) & 4 (-30, 39) &  0.02 (-0.12, 0.16) & 1 \\ 
			60-69 & female & 232 & 207 (188, 247) & 25 (-15, 44) &  0.12 (-0.07, 0.21) & 2 \\ 
			60-69 & male & 355 & 361 (307, 416) & -6 (-61, 48) & -0.02 (-0.17, 0.13) & 13 \\ 
			70-79 & female & 507 & 501 (454, 550) & 6 (-43, 53) &  0.01 (-0.09, 0.11) & 13 \\ 
			70-79 & male & 686 & 694 (632, 779) & -8 (-93, 54) & -0.01 (-0.13, 0.08) & 21 \\ 
			80+ & female & 2217 & 2090 (1957, 2225) & 127 (-8, 260) &  0.06 ( 0.00, 0.12) & 54 \\ 
			80+ & male & 1677 & 1497 (1395, 1618) & 180 (59, 282) &  0.12 ( 0.04, 0.19) & 50 \\ 
			0-39 & Total & 132 & 116 (98, 139) & 16 (-7, 34) & 0.14 (-0.06, 0.29) & 0 \\ 
			40-59 & Total & 403 & 390 (353, 429) & 13 (-26, 50) & 0.03 (-0.07, 0.13) & 6 \\ 
			60-69 & Total & 587 & 572 (502, 640) & 15 (-53, 85) & 0.03 (-0.09, 0.15) & 15 \\ 
			70-79 & Total & 1193 & 1195 (1106, 1326) & -2 (-133, 87) & 0.00 (-0.11, 0.07) & 34 \\ 
			80+ & Total & 3894 & 3590 (3415, 3772) & 304 (122, 479) & 0.08 ( 0.03, 0.13) & 104 \\ 
			Total & female & 3163 & 2994 (2842, 3144) & 169 (19, 321) & 0.06 (0.01, 0.11) & 74 \\ 
			Total & male & 3046 & 2875 (2726, 3031) & 171 (15, 320) & 0.06 (0.01, 0.11) & 85 \\ 
			Total & Total & 6209 & 5868 (5648, 6096) & 341 (113, 561) & 0.06 (0.02, 0.10) & 159 \\ 
			\hline
		\end{tabular}
	\end{table}
	
	\clearpage

	\begin{table}[ht]
		\caption{Expected, excess, relative excess and COVID-19 mortality during \textbf{August} 2022 in Switzerland by age and sex. For the estimated of expected, excess and relative excess mortality we show median and 95\% credible intervals (The probability that the truth lies inside this interval is 0.95).}
		\centering
		\begin{tabular}{llrrrrrr}
			\hline
			Age & Sex & Deaths & Expected & Excess & Relative Excess & COVID-19 \\
			\hline
			0-39 & female & 47 & 40 (28, 59) & 7 (-12, 19) & 0.17 (-0.30, 0.48) & 0 \\ 
			0-39 & male & 76 & 75 (57, 95) & 1 (-19, 19) & 0.01 (-0.25, 0.25) & 0 \\ 
			40-59 & female & 150 & 148 (124, 182) & 2 (-32, 26) & 0.01 (-0.22, 0.18) & 0 \\ 
			40-59 & male & 235 & 235 (200, 281) & 0 (-46, 35) & 0.00 (-0.20, 0.15) & 0 \\ 
			60-69 & female & 237 & 214 (186, 239) & 23 (-2, 51) & 0.11 (-0.01, 0.24) & 4 \\ 
			60-69 & male & 355 & 350 (302, 413) & 5 (-58, 53) & 0.01 (-0.17, 0.15) & 1 \\ 
			70-79 & female & 491 & 492 (443, 553) & -1 (-62, 48) & 0.00 (-0.13, 0.10) & 9 \\ 
			70-79 & male & 681 & 678 (612, 764) & 3 (-83, 69) & 0.00 (-0.12, 0.10) & 11 \\ 
			80+ & female & 2103 & 2053 (1902, 2193) & 50 (-90, 201) & 0.02 (-0.04, 0.10) & 24 \\ 
			80+ & male & 1476 & 1450 (1343, 1574) & 26 (-98, 133) & 0.02 (-0.07, 0.09) & 25 \\ 
			0-39 & Total & 123 & 115 (88, 145) & 8 (-22, 35) & 0.07 (-0.19, 0.30) & 0 \\ 
			40-59 & Total & 385 & 383 (338, 445) & 2 (-60, 47) & 0.01 (-0.16, 0.12) & 0 \\ 
			60-69 & Total & 592 & 566 (511, 629) & 26 (-37, 81) & 0.05 (-0.07, 0.14) & 5 \\ 
			70-79 & Total & 1172 & 1173 (1079, 1300) & -1 (-128, 93) & 0.00 (-0.11, 0.08) & 20 \\ 
			80+ & Total & 3579 & 3503 (3313, 3693) & 76 (-114, 266) & 0.02 (-0.03, 0.08) & 49 \\ 
			Total & female & 3028 & 2948 (2780, 3110) & 80 (-82, 248) & 0.03 (-0.03, 0.08) & 37 \\ 
			Total & male & 2823 & 2798 (2644, 2970) & 25 (-147, 179) & 0.01 (-0.05, 0.06) & 37 \\ 
			Total & Total & 5851 & 5745 (5516, 5992) & 106 (-141, 335) &  0.02 (-0.02,  0.06) & 74 \\ 
			\hline
		\end{tabular}
	\end{table}
	
	\clearpage
	
	\begin{table}[ht]
		\caption{Median and 95\% credible intervals of the relative heatwave all-cause mortality risk and number of deaths attributable to heatwaves (defined as at least 3 consecutive days of temperatures higher than the 90\% temperature percentile and a 3-day lag accounting for any prolonged effects of the heatwaves). Results are shown for the different age and sex groups together with the number of deaths during a heatwave for the corresponding subgroups in summer 2022. Results are adjusted for COVID-19 mortality, cantonal holidays, day of week, spatiotemporal trends and presented with and without ambient temperature adjustment.}
		\label{tab3}
		\centering
		\begin{tabular}{llrrrrrr}
			\hline
			&  &  & \multicolumn{2}{c}{Model with temperature} & \multicolumn{2}{c}{Model without Temperature} & \\
			\hline
			Age & Sex & Deaths & Effect & Deaths & Effect & Deaths & Excess \\ 
			\hline
			0-39 & male & 36.00 & 0.98 (0.84, 1.16) & -1 (-7, 5) & 1.02 (0.89, 1.17) & 1 (-4, 5) & -3 (-18, 6) \\ 
			40-59 & male & 125.00 & 0.97 (0.89, 1.05) & -4 (-16, 6) & 1.01 (0.94, 1.09) & 1 (-9, 10) & 0 (-21, 24) \\ 
			60-69 & male & 157.00 & 0.94 (0.87, 1.02) & -10 (-24, 2) & 0.97 (0.91, 1.04) & -5 (-16, 6) & -25 (-52, 1) \\ 
			70-79 & male & 365.00 & 1.00 (0.94, 1.06) & -1 (-24, 21) & 1.02 (0.96, 1.08) & 6 (-14, 26) & 12 (-41, 50) \\ 
			80+ & male & 918.00 & 1.06 (1.01, 1.11) & 49 (9, 89) & 1.11 (1.07, 1.16) & 91 (59, 125) & 115 (37, 184) \\ 
			0-39 & female & 17.00 & 0.87 (0.70, 1.09) & -3 (-7, 1) & 1.02 (0.84, 1.23) & 0 (-3, 3) & -5 (-17, 3) \\ 
			40-59 & female & 77.00 & 0.90 (0.80, 1.02) & -8 (-19, 2) & 1.00 (0.90, 1.11) & 0 (-8, 8) & -5 (-26, 9) \\ 
			60-69 & female & 113.00 & 0.92 (0.84, 1.02) & -9 (-21, 2) & 0.96 (0.88, 1.04) & -5 (-15, 4) & 2 (-19, 25) \\ 
			70-79 & female & 247.00 & 1.00 (0.94, 1.07) & 0 (-16, 16) & 1.04 (0.98, 1.10) & 9 (-5, 23) & -14 (-52, 12) \\ 
			80+ & female & 1175.00 & 1.06 (1.02, 1.09) & 63 (22, 102) & 1.10 (1.07, 1.14) & 107 (73, 140) & 36 (-51, 126) \\ 
			\hline
		\end{tabular}
	\end{table}
	
	\clearpage
	
	\begin{table}[ht]
		\caption{Median and 95\% credible intervals of the relative heatwave COVID-19 mortality risk and number of COVID-19 deaths attributable to heatwaves (defined as at least 3 consecutive days of temperatures higher than the 90\% temperature percentile and a 3-day lag accounting for any prolonged effects of the heatwaves). Results are shown for the different age and sex groups together with the number of COVID-19 deaths during a heatwave and total number of COVID-19 deaths during summers 2021-2022. Results are adjusted for cantonal holidays, day of week, spatiotemporal trends and presented with and without ambient temperature adjustment.}
		\label{tab4}
		\centering
		\begin{tabular}{llrrrr}
			\hline
			&  & \multicolumn{2}{c}{COVID-19 Deaths} & \multicolumn{2}{c}{Model}  \\
			\hline
			age & sex & Heatwaves & Total & Effect & Deaths \\ 
			\hline
			0-39 & male & 0 & 2 &  - & - \\ 
			40-59 & male & 2 & 9 & 3.19 (0.48, 21.83) & 1 (-1, 1) \\ 
			60-69 & male & 2 & 34 & 0.22 (0.05, 1.04) & -4 (-19, 0) \\ 
			70-79 & male & 13 & 74 & 0.89 (0.43, 1.84) & -1 (-9, 3) \\ 
			80+ & male & 31 & 135 & 1.43 (0.84, 2.45) & 7 (-4, 13) \\ 
			0-39 & female & 0 & 0 & - & - \\ 
			40-59 & female & 3 & 19 & 2.46 (0.16, 14.14) & 1 (-10, 2) \\ 
			60-69 & female & 2 & 18 & 0.64 (0.03, 4.44) & 0 (0, 0) \\ 
			70-79 & female & 9 & 39 & 1.29 (0.25, 5.76) & 2 (-21, 6) \\ 
			80+ & female & 42 & 141 & 1.84 (1.16, 2.90) & 15 (5, 22) \\ 
			\hline
		\end{tabular}
	\end{table}
	
	\clearpage

	\begin{table}[ht]
		\caption{Expected, excess, relative excess and COVID-19 mortality during summer 2022 in Switzerland by age and sex for the model \textbf{without accounting for ambient temperature}. For the estimated of expected, excess and relative excess mortality we show median and 95\% credible intervals (The probability that the truth lies inside this interval is 0.95).}
		\centering
		\begin{tabular}{llrrrrrr}
			\hline
			Age & Sex & Deaths & Expected & Excess & Relative Excess & COVID-19 \\
			\hline
			0-39 & female & 126 & 116 (92, 153) & 10 (-27, 34) & 0.09 (-0.23, 0.29) & 0 \\ 
			0-39 & male & 235 & 221 (192, 257) & 14 (-22, 43) & 0.06 (-0.10, 0.19) & 0 \\ 
			40-59 & female & 444 & 435 (383, 509) & 9 (-65, 61) & 0.02 (-0.15, 0.14) & 7 \\ 
			40-59 & male & 709 & 699 (637, 776) & 10 (-67, 72) & 0.01 (-0.10, 0.10) & 1 \\ 
			60-69 & female & 657 & 624 (579, 692) & 33 (-35, 78) & 0.05 (-0.06, 0.12) & 7 \\ 
			60-69 & male & 1070 & 1055 (945, 1180) & 15 (-110, 125) & 0.01 (-0.10, 0.12) & 16 \\ 
			70-79 & female & 1460 & 1455 (1360, 1573) & 5 (-113, 100) & 0.00 (-0.08, 0.07) & 24 \\ 
			70-79 & male & 2047 & 2019 (1872, 2228) & 28 (-181, 175) & 0.01 (-0.09, 0.09) & 44 \\ 
			80+ & female & 6326 & 6007 (5692, 6396) & 319 (-70, 634) & 0.05 (-0.01, 0.11) & 97 \\ 
			80+ & male & 4528 & 4280 (4048, 4627) & 248 (-99, 480) & 0.06 (-0.02, 0.11) & 89 \\ 
			0-39 & Total & 361 & 337 (296, 384) & 24 (-23, 65) & 0.07 (-0.07, 0.19) & 0 \\ 
			40-59 & Total & 1153 & 1141 (1057, 1240) & 12 (-87, 96) & 0.01 (-0.08, 0.08) & 8 \\ 
			60-69 & Total & 1727 & 1674 (1567, 1823) & 53 (-96, 160) & 0.03 (-0.06, 0.10) & 23 \\ 
			70-79 & Total & 3507 & 3478 (3298, 3783) & 29 (-276, 209) & 0.01 (-0.08, 0.06) & 68 \\ 
			80+ & Total & 10854 & 10318 (9867, 10832) & 536 (22, 987) & 0.05 ( 0.00, 0.10) & 186 \\ 
			Total & female & 9013 & 8647 (8291, 9074) & 366 (-61, 722) & 0.04 (-0.01, 0.08) & 135 \\ 
			Total & male & 8589 & 8318 (7956, 8736) & 272 (-147, 633) & 0.03 (-0.02, 0.08) & 150 \\ 
			Total & Total & 17602 & 16974 (16440, 17579) & 628 (23, 1162) & 0.04 (0.00, 0.07) & 285 \\ 
			\hline
		\end{tabular}
	\end{table}
	
	\clearpage
	
	\begin{figure}
		\caption{Root mean square error across the different age and sex groups based on the cross validation performed for the 5 population models}
		\centering
		\includegraphics{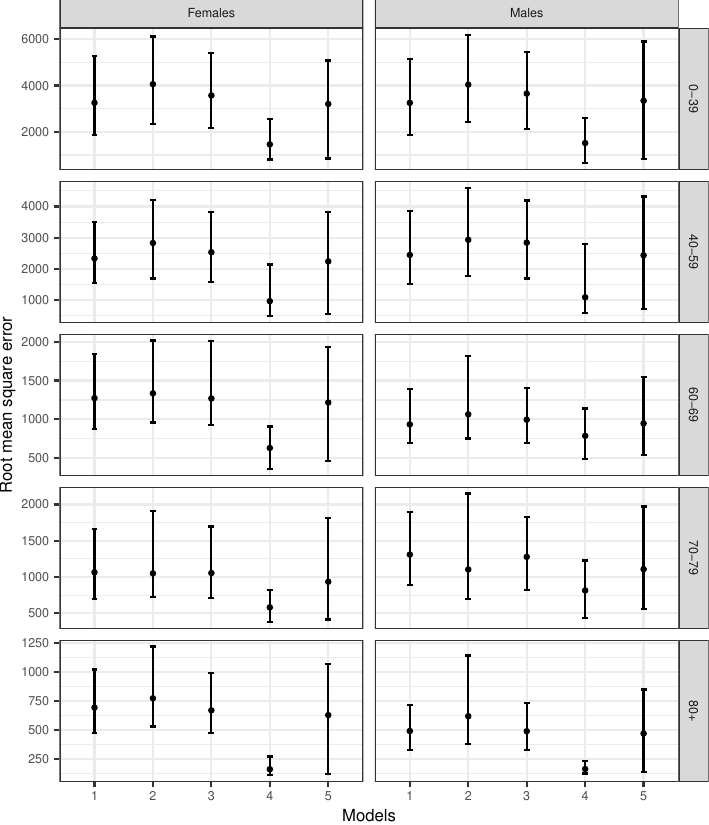}
	\end{figure}
	
	\clearpage
	
	\clearpage
	
	\begin{figure}
		\caption{Bias across the different age and sex groups based on the cross validation performed for the 5 population models. The red line highlights zero.}
		\centering
		\includegraphics{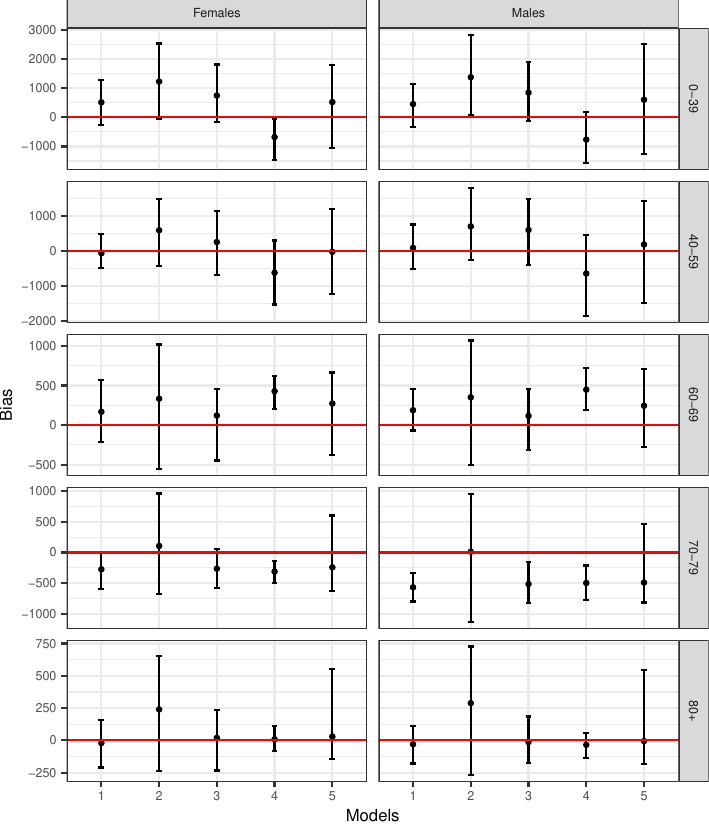}
		
	\end{figure}
	
	\clearpage
	
	\begin{figure}
		\caption{Coverage probability across the different age and sex groups based on the cross validation performed for the 5 population models.}
		\centering
		\includegraphics{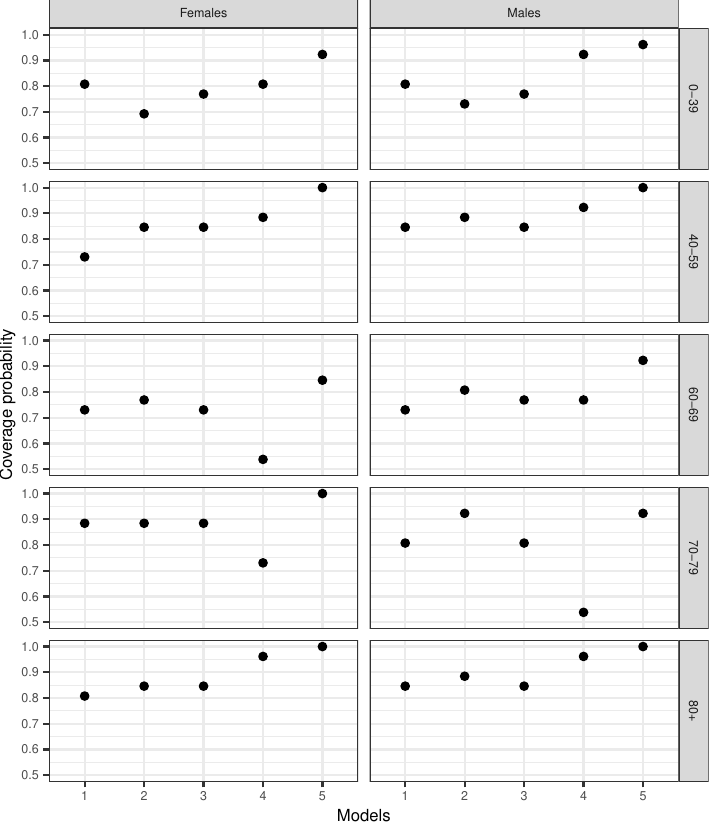}
	\end{figure}
	
	\clearpage
	
	\begin{figure}
		\caption{200 samples of bias [Truth - Predicted number of deaths] across the different age and sex groups based on the cross validation performed for the 12 models used to estimate excess mortality.}
		\centering
		\includegraphics[scale=0.9]{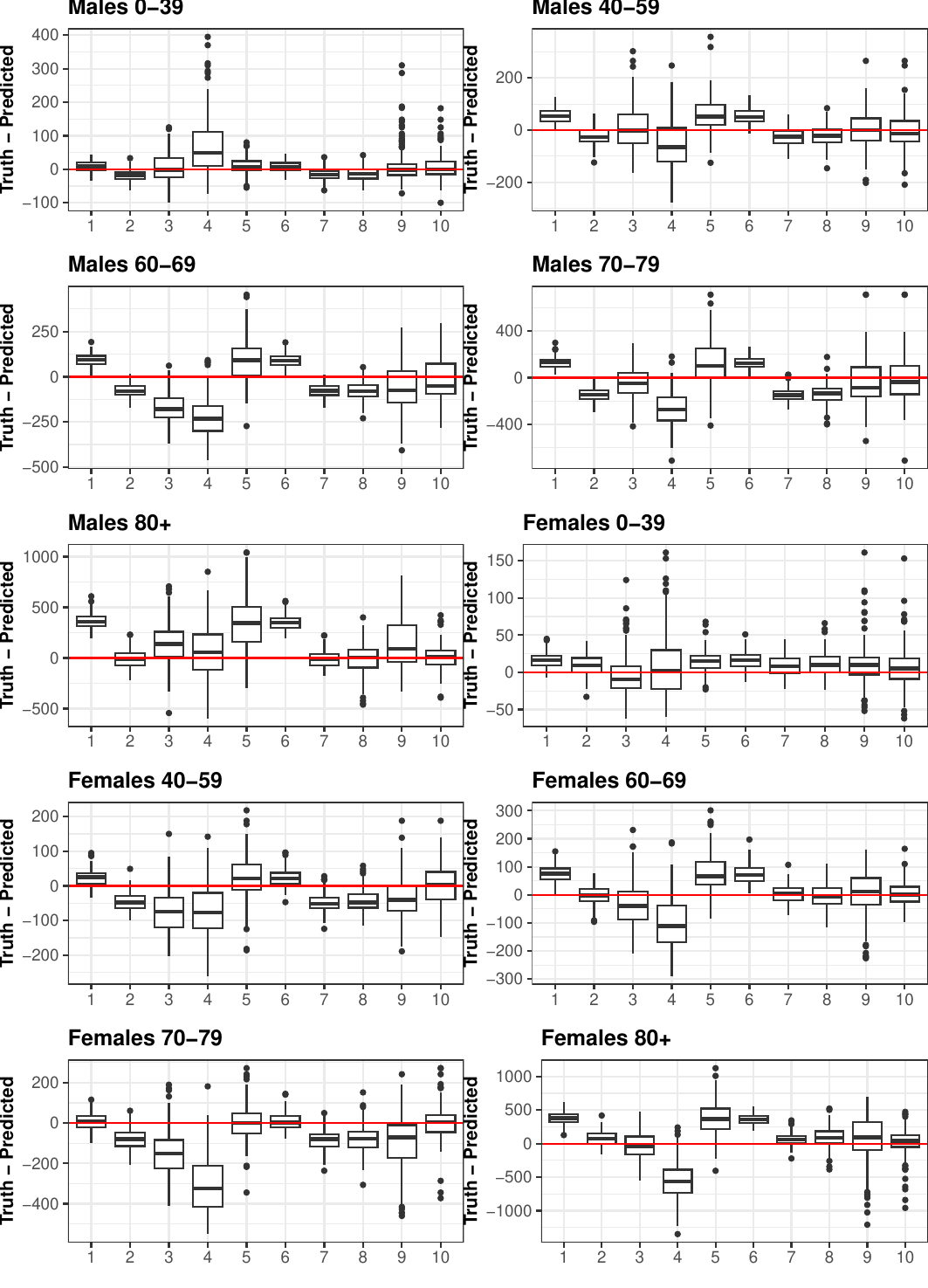}
		
	\end{figure}
	
	\clearpage
	
	\begin{figure}
		\caption{The cantons of Switzerland.}
		\centering
		\includegraphics[scale=0.9]{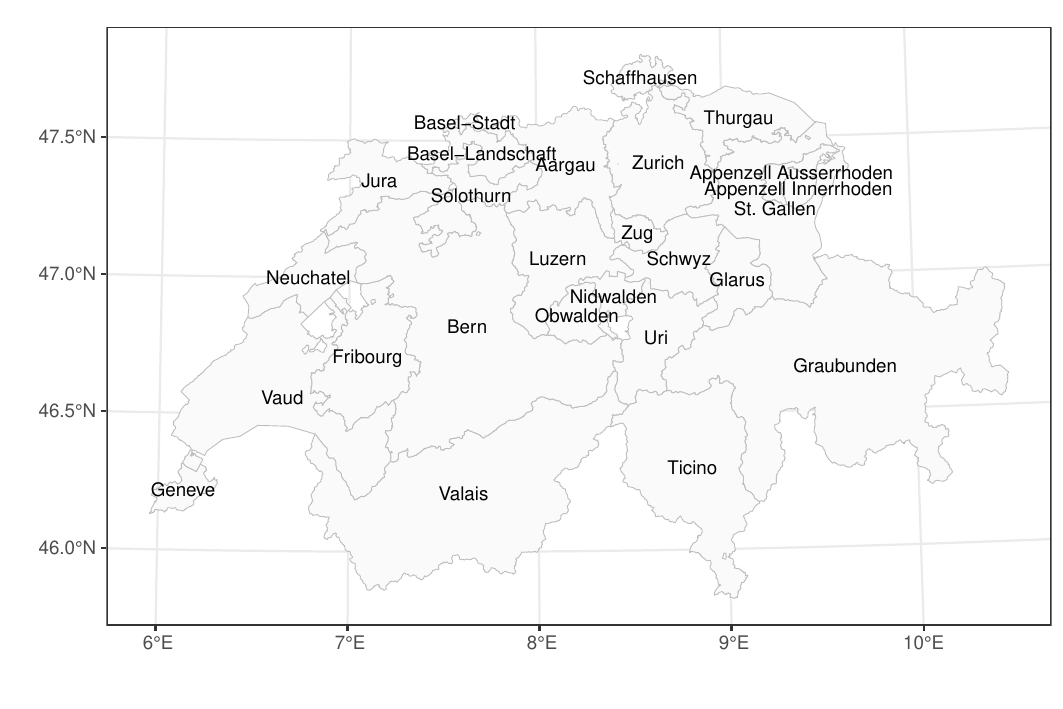}
	\end{figure}
	
	\clearpage

	\begin{figure}
		\caption{Median and 95\% credible intervals of the relative heatwave related mortality risk during 2011-2022 across the different age groups.}
		\centering
		\includegraphics[scale=0.9]{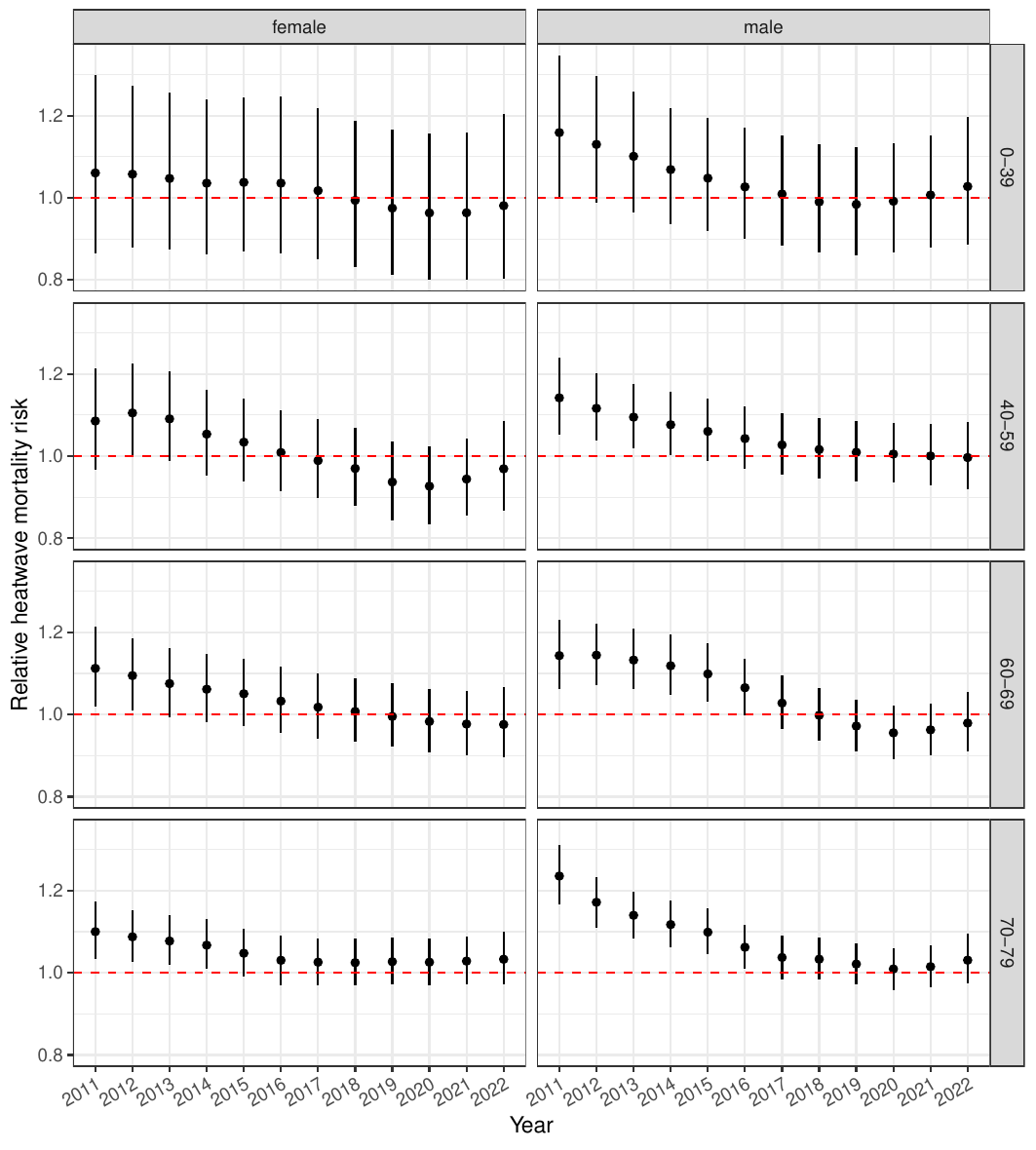}
	\end{figure}
	
	\clearpage
	
	\begin{figure}[ht]
		\caption{Median and 95\% credible intervals of the relative heatwave related mortality risk during 2011-2022 for people aged older than 80 years.}
		\label{fig3}
		\centering
		\includegraphics[scale=0.9]{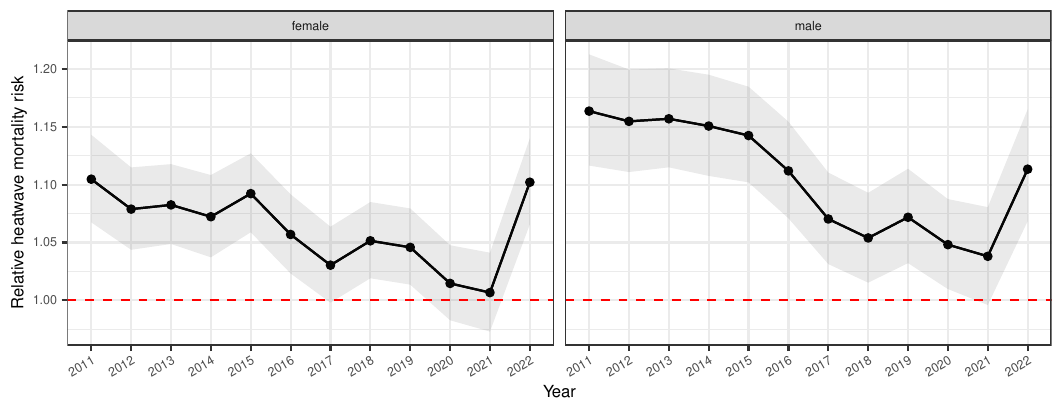}
	\end{figure}
	
	\clearpage
	
	\begin{figure}
		\caption{Number of excess deaths, excess deaths per heatwave and relative excess mortality across the different definitions of a heatwave (regarding duration and temperature thresholds). The blue line on the panel on the left shows a hypothetical scenario where the excess is independent of thresholds that define periods of higher temperature.}
		\centering
		\includegraphics[scale=0.9]{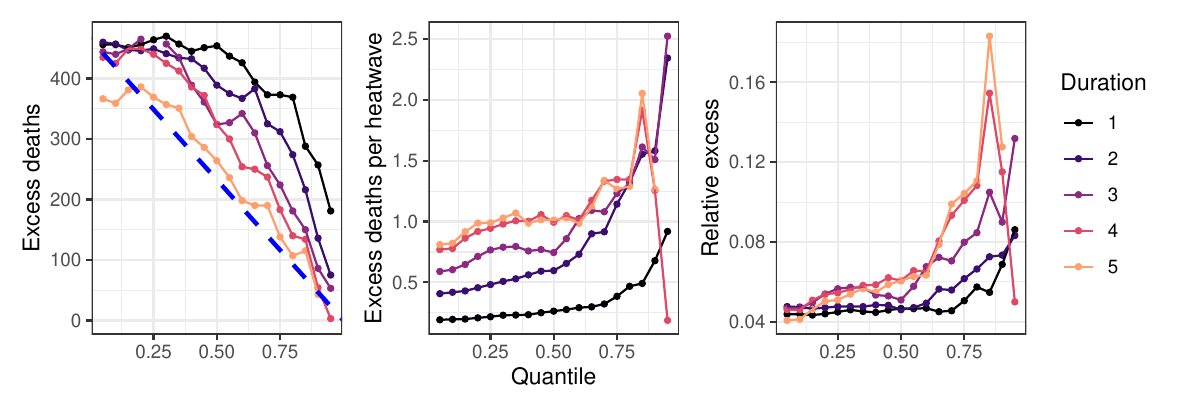}
	\end{figure}

\end{document}